\documentstyle[12pt,ccgrra]{article}
\newcommand{\D}{{\cal{D}}}
\newcommand{\eqref}[1]{(\ref{#1})}
\newcommand{\xt}{\tilde{x}}
\begin{document}

\title{SKELETONIZATIONS OF PHASE SPACE PATHS}

\author{John T. Whelan\footnote{e-mail: whelan@physics.utah.edu}} 

\affil{University of Utah, Department of Physics\\
115 S 1400 E Room 201,  Salt Lake City, UT 84112-0830}

\beginabstract 
Construction of skeletonized path integrals for a particle moving on a
curved spatial manifold is considered.  As shown by DeWitt, Kucha\v{r}
and others, while the skeletonized configuration space action can be
written unambiguously as a sum of Hamilton principal functions,
different choices of the measure will lead to different
Schr\"{o}dinger equations.  On the other hand, the Liouville measure
provides a unique measure for a skeletonized phase space path
integral, but there is a corresponding ambiguity in the
skeletonization of a path through phase space.  A family of
skeletonization rules described by Kucha\v{r} and referred to here as
geodesic interpolation is discussed, and shown to behave poorly under
the involution process, wherein intermediate points are removed by
extremization of the skeletonized action.  A new skeletonization rule,
tangent interpolation, is defined and shown to possess the desired
involution properties.
\endabstract

\section{SKELETONIZED PATH INTEGRALS}\label{sec:skel}

	The path integral method seems to provide a generic recipe for
producing a quantum theory given a classical action.  This recipe is
based on the principle that amplitudes are given by sums, over the
relevant histories, of the complex exponential of the action.  For
example, a wavefunction $\psi(x_i,t_i)$ is propagated to a later time
$t_f>t_i$ by
\begin{equation}\label{formconf}
\psi(x_f,t_f)=\int\limits_{x_f\phantom{x_i}}
\D^n\! x\, e^{iS[x]/\hbar}\psi(x_i,t_i),
\end{equation}
where the integral is over all paths ending at the argument $x_f$.
However, in order to provide a constructive definition for an
expression like \eqref{formconf} one needs to give a meaning to formal
concepts like an integral over paths, and it is in this step that
decisions need to be made in the implementation of the quantum theory.

	The system considered in this work is a free non-relativistic
particle of unit mass moving on a curved $n$-dimensional
\emph{spatial} manifold, which is described by the action
\begin{equation}\label{freepart}
S[x]=\frac{1}{2}\int dt\,g_{ab}(x)\frac{dx^a}{dt}\frac{dx^b}{dt}.
\end{equation}
This is the simplest action which exhibits the ambiguities considered
here, which correspond in an operator theory to questions of operator
ordering.  One can also modify the action to include potential terms,
but previous work \cite{kuchar} suggests that the results will remain
essentially the same.

\subsection{Configuration Space}

	A formal path integral such as \eqref{formconf} can be given a
concrete definition via the skeletonization process, in which the
interval $(t_i,t_f)$ is broken up by a series of $N$ time instants
$\{t_{(I)}|I=1,\ldots N\}$.  The integral over all paths $x(t)$ is
realized, in the limit of an infinitely fine time slicing, as a
product of integrals over the positions $\{x_{(I)}|I=0,\ldots N\}$ of
the particle at the discrete times:
\begin{equation}\label{skelconf}
\psi(x_f,t_f)=\lim_{N\rightarrow\infty}
\left(\prod_{I=0}^N\int A(x_{(I+1)}t_{(I+1)}|x_{(I)} t_{(I)})\,d^n\! x_{(I)}\,
e^{iS(x_{(I+1)} t_{(I+1)}|x_{(I)} t_{(I)})/\hbar}\right)\psi(x_i,t_i),
\end{equation}
where $t_{(0)}=t_i$ and $t_{(N+1)}=t_f$.  This process replaces the
action functional $S[x]$ with a function $\sum_{I=0}^N S(x_{(I+1)}
t_{(I+1)}|x_{(I)} t_{(I)})$ which is constructed by summing
contributions from the intervals between time slices.  Since the
equations of motion are second order, there is a preferred path
between a pair of endpoints $(x,t)$ and $(x',t')$, namely, the
classical path between them, which for the theory described by the
action \eqref{freepart} is an affinely parametrized
geodesic.\footnote{Throughout this paper, the points in question, even
when separated by a finite distance, are taken to be close enough that
there is no crossing of geodesics.}  The action functional evaluated
on that piece of the path is the \emph{Hamilton principal function}
$S(x't'|x\,t)$.  On the other hand, the measure $\prod_{I=0}^N\int
A(x_{(I+1)}t_{(I+1)}|x_{(I)} t_{(I)})d^n\! x_{(I)}$ for the path
integral has no such natural definition in curved space.  Feynman
\cite{feyn} showed that $A(x't'|x\,t)=[2\pi i\hbar(t'-t)]^{-n/2}$ was
the appropriate measure for flat space in Cartesian co\"{o}rdinates.
The simplest generalization of the Feynman measure to curved space is
\cite{dewitt}
\begin{equation}\label{feynmeas}
A_0(x't'|x\,t)=\frac{\sqrt{g}}{[2\pi i\hbar(t'-t)]^{n/2}},
\end{equation}
where $g$ is the determinant of the metric $g_{ab}$ at the point $x$,
which ensures that the path integral is invariant under co\"{o}rdinate
changes.  However, there are other measures which are similarly
invariant and also reduce to the Feynman measure when the spatial
manifold is flat.  Expanding to second order in powers of the geodesic
distance $\sigma'$ between $x$ and $x'$ (which turns out to be
sufficient to evaluate the path integral), one family of such measures
is\footnote{Specific measures of this sort were discussed in
\cite{dewitt,meas}, and placed into the family \eqref{measlamb} in
\cite{kuchar}.}
\begin{equation}\label{measlamb}
A_\lambda(x't'|x\,t)=\frac{\sqrt{g}}{[2\pi i\hbar(t'-t)]^{n/2}}
\left(1+\frac{\lambda}{3} R^{ab} y'_a y'_b+\Or(\sigma'^3)\right),
\end{equation}
where $y'_a$ are the Riemann normal co\"{o}rdinates of the point $x'$
with respect to the origin $x$ and $R^{ab}$ is the Ricci tensor at
$x$.

	This ambiguity in defining the path integral measure has
tangible consequences.  Namely, if we use a member of the family
\eqref{measlamb} to define the measure in the skeletonized path
integral \eqref{skelconf}, the wavefunction thus propagated satisfies
\cite{dewitt,kuchar} a Schr\"{o}dinger equation
\begin{equation}
i\hbar\frac{\partial\psi(x,t)}{\partial t}=
\left(-\frac{\hbar^2}{2}\,g^{ab}\nabla_a\nabla_b
+\frac{1-\lambda}{6}\hbar^2 R\right)
\psi(x,t).
\end{equation}
The classical Hamiltonian $g^{ab}p_a p_b$ is not quantized in the
simplest way, by replacing it with an operator proportional to the
covariant Laplacian, but also contains a measure-dependent term
proportional to the scalar curvature.  In particular, the
straightforward generalization \eqref{feynmeas} of the Feynman measure
(in which $\lambda=0$) leads to a Schr\"{o}dinger equation \emph{with}
a curvature correction term.

\subsection{Phase Space}

	One way to resolve the measure ambiguity is to start instead
with a phase space path integral \cite{garrod}
\begin{equation}\label{formphas}
\psi(x_f,t_f)=\int\limits_{x_f\phantom{x_i}}
\D^n\! p\,\D^n\! x\, e^{iS[p,x]/\hbar}\psi(x_i,t_i),
\end{equation}
in which the path integral measure is expressed as a product of
Liouville measures $d^n\!p\,d^n\!x/(2\pi\hbar)^n$ in the
skeletonization
\begin{equation}\label{skelphas}
\psi(x_f,t_f)=\lim_{N\rightarrow\infty}
\left(\prod_{I=0}^N\int \frac{d^n\! p^{(I)}\,d^n\! x_{(I)}}{(2\pi\hbar)^n}
e^{iS(x_{(I+1)} t_{(I+1)}|p^{(I)} x_{(I)} t_{(I)})/\hbar}\right)\psi(x_i,t_i).
\end{equation}
Unfortunately, the ambiguity has now shifted into the definition of
the contributions $\{S(x_{(I+1)} t_{(I+1)}|p^{(I)} x_{(I)} t_{(I)})\}$
to the skeletonized action.  This is because there is in general no
classical trajectory through phase space satisfying the conditions
\begin{equation}\label{BCs}
x(t)=x,\qquad p(t)=p, \qquad x(t')=x',
\end{equation}
since any two of these three conditions are sufficient to specify a
classical phase space path.

	Since there is no \emph{classical} phase space path, we need
some interpolation rule which associates a \emph{virtual} path through
phase space with the set of boundary conditions \eqref{BCs}.  (It
should of course reproduce the classical path when the three boundary
conditions are classically consistent.)  A given interpolation rule
defines a \emph{phase space principal function} (PSPF)
$S(x't'|p\,x\,t)$ whose value for a particular set of arguments will
be the action functional evaluated along the virtual path which the
rule specifies for those boundary conditions.  The interpolation rule
then defines a skeletonization, under which and the skeletonized
canonical action appearing in \eqref{skelphas} is a sum of PSPFs.

\subsubsection{Geodesic Interpolation.}
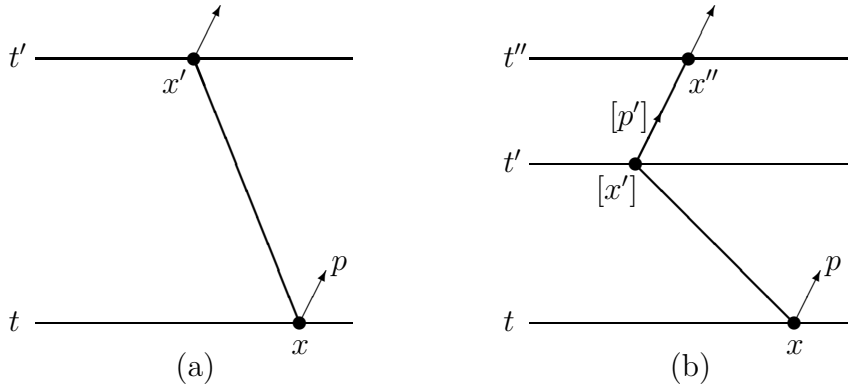
\begin{figure}
$$
\begin{picture}(140,140)(-10,-20)
\put(53,-20){(a)}
\put(0,0){\line(1,0){120}}
\put(-10,-3){$t$}
\put(0,100){\line(1,0){120}}
\put(-10,97){$t'$}
\put(100,0){\circle*{5}}
\put(97,-12){$x$}
\put(100,0){\vector(1,2){10}}
\put(112,20){$p$}
\put(60,100){\circle*{5}}
\put(48,87){$x'$}
\put(60,100){\vector(1,2){10}}
\thicklines
\put(100,0){\line(-2,5){40}}
\end{picture}
\qquad\qquad
\begin{picture}(140,140)(-10,-20)
\put(53,-20){(b)}
\put(0,0){\line(1,0){120}}
\put(-10,-3){$t$}
\put(0,60){\line(1,0){120}}
\put(-10,57){$t'$}
\put(0,100){\line(1,0){120}}
\put(-10,97){$t''$}
\put(100,0){\circle*{5}}
\put(97,-12){$x$}
\put(100,0){\vector(1,2){10}}
\put(112,20){$p$}
\put(40,60){\circle*{5}}
\put(25,47){$[x']$}
\put(40,60){\vector(1,2){10}}
\put(30,75){$[p']$}
\put(60,100){\circle*{5}}
\put(60,87){$x''$}
\put(60,100){\vector(1,2){10}}
\thicklines
\put(100,0){\line(-1,1){60}}
\put(40,60){\line(1,2){20}}
\end{picture}
$$
\caption{Geodesic interpolation (GI), illustrated in $1+1$ dimensions,
with the time direction running vertically and the space direction
running horizontally.  \textbf{(a)} Constructing a phase space path
from the boundary conditions (\protect\ref{BCs}); the configuration
space trajectory is the classical path, a straight line from $x$ to
$x'$, while the momentum $p$ is held constant along the trajectory.
In higher dimensions, if the spatial manifold is curved, the GI scheme
also requires a rule for propagation of the momentum.  \textbf{(b)}
Failure of involution of paths for GI; using the values
(\protect\ref{flatinvo}) of $x'$ and $p'$ which extremize the sum
(\protect\ref{flatpfs}) of phase space principal functions, one
constructs the two-step phase space trajectory, with the first step
defined by $x$, $p$ and $x'$ and the second by $x'$, $p'$ and $x''$.
Comparison to Fig.~\protect\ref{fig:GI}(a) shows that this is not the
same path as would be constructed from $x$, $p$ and $x''$ by the GI
prescription (although the two paths do have the same action when the
space is flat).}
\label{fig:GI}
\end{figure}

	One type of rule, introduced by Kucha\v{r} \cite{kuchar},
prescribes the configuration space path to be the classical path from
$x$ to $x'$, regardless of the value of $p$.  The momentum then ``goes
along for the ride,'' being propagated in some linear way along the
geodesic (Fig.~\ref{fig:GI}a).
The resulting phase space principal function is
\begin{equation}\label{geodpspf}
S(x't'|p\,x\,t)=p^a y'_a-\frac{t'-t}{2}G^{ab}(x'|x)p_a p_b\, ,
\end{equation}
with the functional form of $G^{ab}(x'|x)$ depending on the details of
the propagation rule.  We call this ``geodesic interpolation'' (GI)
because the property of the classical trajectories which is retained
by all of the virtual ones is that the configuration space projection
of the path is a geodesic.  This motivation for selecting rules in
which the second order equation of motion, involving only the path
$x(t)$, is obeyed independent of the path $p(t)$, comes from the
invariance of the tensor formulation under point canonical
transformations [$ x\rightarrow X(x), p\rightarrow P(p,x) $], in which
the transformation of the $x$ components of a point in phase space is
independent of its $p$ components, but not under general canonical
transformations [$x\rightarrow X(p,x), p\rightarrow P(p,x)$].

	If the momentum $p$ is taken to be parallel-propagated along
the geodesic, the tensor $G^{ab}(x'|x)$ in \eqref{geodpspf} becomes
the inverse metric $g^{ab}$ at $x$.  This is one of a family of
propagation rules discussed by Kucha\v{r} which, for small intervals,
have the expansion
\begin{equation}\label{Glambda}
G_\lambda^{ab}(x'|x)=g^{ab}-\frac{\lambda}{3} R^{acbd} y'_c y'_d
+\Or(\sigma'^3),
\end{equation}
where $R^{abcd}$ is the Riemann curvature tensor at $x$.

	Performing the quadratic integrals over the momenta in the
skeletonized path integral \eqref{skelphas} induces a measure
\begin{equation}
A(x't'|x\,t)=\frac{1}{\sqrt{[2\pi i\hbar(t'-t)]^n\det G^{ab}}}
\end{equation}
on the remaining configuration space path integrals.  The family of
propagation rules given by \eqref{Glambda} lead in this way to exactly
the family of configuration space measures \eqref{measlamb}.  This
means that they produce quantum theories whose Hamiltonians contain
curvature correction terms $\frac{1-\lambda}{6}\hbar^2 R$; in
particular, the parallel propagation rule ($\lambda=0$) leads to the
Feynman measure \eqref{feynmeas} and hence to a curvature-corrected
quantum Hamiltonian.  The case $\lambda=1$, which Kucha\v{r} showed
arose from a propagation rule based on the equation of geodesic
deviation, leads to a Hamiltonian with no curvature correction term.

	This work considers a criterion for selecting a phase space
skeletonization rule, and hence a principal function.  This criterion,
as described in the following section, is based on a property of the
PSPFs and paths themselves.

\section{THE INVOLUTION PROPERTY}\label{sec:invo}

\subsection{Configuration space}

	We will judge phase space skeletonizations by whether they
obey a property analogous to the involution property satisfied by
configuration space skeletonizations.  The configuration space
property is defined as follows: Consider the contribution to the
skeletonized action from the intervals between three consecutive time
instants, called $t$,~$t'$ and~$t''$:
\begin{equation}\label{compconf}
S(x''t''|x't')+S(x't'|x\,t).
\end{equation}
It is a function of three positions, $x$, $x'$ and $x''$, and has
associated with it a path made up of two geodesics connected at the
point $x'$.  If we vary the location of that point $x'$ while holding
the endpoints $x$ and $x''$ fixed, the expression \eqref{compconf} is
minimized by the $x'$ which lies at the appropriate point on the
geodesic from $x$ to $x''$.  The path corresponding to that choice of
$x'$ is just the geodesic connecting the two endpoints $x$ and $x''$,
and its action is the Hamilton principal function $S(x''t''|x\,t)$:
\begin{equation}\label{invoconf}
S(x''t''|x't')+S(x't'|x\,t)\mathrel{\mathop{\Longrightarrow}^{x'}}
S(x''t''|x\,t).
\end{equation}
This property is known as involution.

	We say that the configuration space skeletonization procedure
obeys
\emph{involution of paths} (IOP) because extremization of
\eqref{compconf} with respect to the intermediate position $x'$ leads
to the classical path from $x$ and $x''$, and
\emph{involution of functions} (IOF) because \eqref{invoconf} holds.
Clearly, 
the latter property is implied by the former.

\subsection{Phase space}

	Given an interpolation rule, which defines a phase space
principal function $S(x't'|p\,x\,t)$, we can define the analogous
property for a \emph{phase space} skeletonization by considering two
consecutive intervals in the skeletonization.  The five quantities
$x$, $p$, $x'$, $p'$ and $x''$ determine a phase space path between
times $t$ and $t''$ according to the interpolation rule.  The action
for this path is
\begin{equation}\label{compphas}
S(x''t''|p'x't')+S(x't'|p\,x\,t).
\end{equation}
Extremization with respect to the intermediate variables $x'$ and $p'$
leaves a function of $x$, $p$ and $x''$; we say that involution of
\emph{functions} holds if that is the same PSPF with which we started:
\begin{equation}\label{invophas}
S(x''t''|p'x't')+S(x't'|p\,x\,t)
\mathrel{\mathop{\Longrightarrow}^{p',x'}}S(x''t''|p\,x\,t).
\end{equation}
The stricter condition of involution of \emph{paths} is satisfied if
the phase space path defined by the extremizing values of $x'$ and
$p'$ is the same one which the interpolation rule would have
constructed \textit{a priori} from $x$, $p$ and $x''$ without making
reference to the intermediate time slice $t'$.

	These concepts are easily illustrated in the case where the
spatial manifold is flat; in that case the preferred propagation rule
associated with geodesic interpolation is simply to keep the momentum
constant (as measured in a Cartesian coordinate system).  This leads
to a PSPF
\begin{equation}\label{flatpspf}
S(x't'|p\,x\,t)=p\cdot(x'-x)-\frac{t'-t}{2}p^2\, .
\end{equation}
The two-step action
\begin{equation}\label{flatpfs}
S(x''t''|p'x't')+S(x't'|p\,x\,t)=p'\cdot x''+(p-p')\cdot x'-p\cdot x
-\frac{t''-t'}{2}p'^2-\frac{t'-t}{2}p^2
\end{equation}
is minimized by
\begin{equation}\label{flatinvo}
p'=p\quad\hbox{and}\quad x'=x''-(t''-t')p.
\end{equation}
Substituting these into \eqref{flatpfs} gives
$p\cdot(x''-x)-\frac{t''-t}{2}p^2=S(x''t''|p\,x\,t)$, so the GI scheme
in flat-space \emph{does} satisfy IOF.  On the other hand, it is easy
to see that it does \emph{not} satisfy IOP, since the path obtained by
extremization has a configuration space projection which starts at $x$
at time $t$, follows a straight path to $x''-(t''-t')p$ at $t'$, and
then makes a sharp turn before heading on to $x''$ at $t''$
(Fig.~\ref{fig:GI}b).  This is not the same as the path from $x$ to
$x''$ which geodesic interpolation would dictate without the presence
of the intermediate point, since that path is just a straight line
from $x$ to $x''$.

	If we look at the curved-space case in the limit that all of
the distances are small, the zeroth-order results of course replicate
the flat-space ones (including the failure of IOP).  If we consider
the family of PSPFs defined by \eqref{Glambda}, the behavior of the
first correction terms will depend on the value of $\lambda$.  It
turns out that for $\lambda=-1$ (and for no other value) IOF continues
to hold to the lowest non-trivial order.

	The reason why IOF can hold even when IOP fails is this: Since
a given PSPF $S(x't'|p\,x\,t)$ corresponds to the action functional
evaluated along a virtual, rather than classical, phase space path
satisfying the boundary conditions \eqref{BCs}, there are many
different paths which yield the same function.  We can use this fact
to our advantage by defining a different skeletonization procedure
which produces the same PSPF over small intervals as $\lambda=-1$
geodesic interpolation, and thus also obeys the IOF property, but
arises from a different family of paths, which satisfies IOP.

\section{TANGENT INTERPOLATION}\label{sec:TI}

	We will call the skeletonization procedure which respects IOP
\emph{tangent interpolation} (TI).  As opposed to the geodesic
interpolation schemes, which require the configuration space part of
the path to be the classical geodesic, the configuration space path in
the TI scheme is not even continuous.  However, the classical equation
relating momentum to velocity is respected by the prescribed path, so
that the momentum vector remains tangent to the configuration space
trajectory.  Hence the name ``tangent interpolation.''

\begin{figure}
$$
\begin{picture}(140,140)(-10,-20)
\put(53,-20){(a)}
\put(0,0){\line(1,0){120}}
\put(-10,-3){$t$}
\put(0,100){\line(1,0){120}}
\put(-10,97){$t'$}
\put(100,0){\circle*{5}}
\put(97,-12){$x$}
\put(100,0){\vector(1,2){10}}
\put(112,20){$p$}
\put(10,0){\circle{5}}
\put(7,-12){$[\xt]$}
\put(10,0){\vector(1,2){10}}
\put(60,100){\circle*{5}}
\put(60,87){$x'$}
\put(60,100){\vector(1,2){10}}
\thicklines
\put(100,0){\line(-1,0){90}}
\put(10,0){\line(1,2){50}}
\end{picture}
\qquad\qquad
\begin{picture}(140,140)(-10,-20)
\put(53,-20){(b)}
\put(0,0){\line(1,0){120}}
\put(-10,-3){$t$}
\put(0,60){\line(1,0){120}}
\put(-10,57){$t'$}
\put(0,100){\line(1,0){120}}
\put(-10,97){$t''$}
\put(100,0){\circle*{5}}
\put(97,-12){$x$}
\put(100,0){\vector(1,2){10}}
\put(112,20){$p$}
\put(40,60){\circle*{5}}
\put(40,47){$[x'=\xt']$}
\put(40,60){\vector(1,2){10}}
\put(30,75){$[p']$}
\put(60,100){\circle*{5}}
\put(60,87){$x''$}
\put(60,100){\vector(1,2){10}}
\put(10,0){\circle{5}}
\put(7,-12){$[\xt]$}
\put(10,0){\vector(1,2){10}}
\thicklines
\put(100,0){\line(-1,0){90}}
\put(10,0){\line(1,2){50}}
\end{picture}
$$
\caption{Tangent interpolation, illustrated in $1+1$ dimensions.
\textbf{(a)} The phase space path associated with the boundary
conditions (\protect\ref{BCs}) has an initial configuration space jump
from $x$ to $\xt$ and then follows the classical phase space trajectory
from $\xt$ to $x'$; the point $\xt$ is chosen so that the prescribed
momentum $p$ agrees with the initial momentum on the geodesic $\xt
x'$.  (In higher-dimensional curved space, this comparison is made in
Riemann normal co\"{o}rdinates based at $x$.)  \textbf{(b)} Involution
of paths for TI.  Using the $x'$ and $p'$ (\protect\ref{flatinvo})
which extremize the sum of phase space principal functions
(\protect\ref{flatpfs}), one constructs a two-step phase space
trajectory, first from $t$ to $t'$ and then from $t'$ to $t''$.  There
is no discontinuity at $t'$ for this choice of $x'$ and $p'$
(\textit{i.e.}, $\xt'=x'$), and the initial jump from $x$ to $\xt$
implied by the chosen $x'$ at time $t'$ is the same as the one implied
by $x''$ at $t''$ (along with the initial $p$ in either case).  Thus
the two-step phase space path is the same as we would get by TI in one
step from $t$ to $t''$.  (Compare Fig.~\ref{fig:TI}a.)}
\label{fig:TI}
\end{figure}
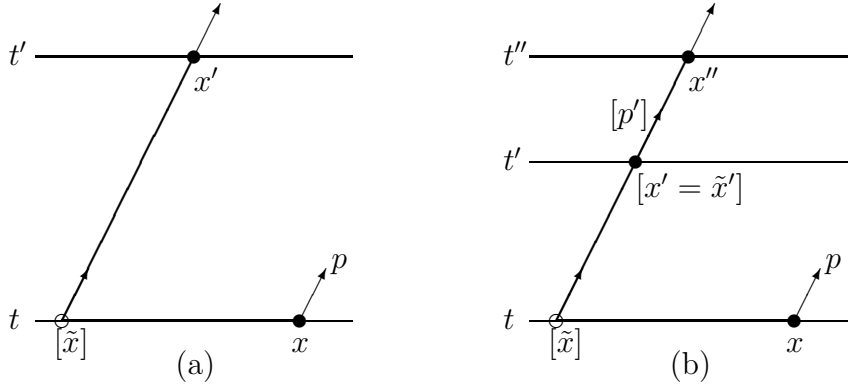

	The phase space path (Fig.~\ref{fig:TI}a)
determined by a given set of boundary conditions \eqref{BCs} begins at
$x$ with the momentum $p$ at time $t$, jumps instantaneously to a
point $\xt$ which is prescribed (in a manner detailed below) by the
boundary conditions, and then follows the classical phase space
trajectory from $\xt$ at $t$ to $x'$ at $t'$.  The point $\xt$ is
chosen so that in a Riemann normal co\"{o}rdinate (RNC) system based
at $x$, there is no momentum discontinuity at the initial time $t$,
\textit{i.e.}, so that the classical momentum at the beginning of the
geodesic from $\xt$ to $x'$ has the same covariant RNC components as
$p$.  In flat space, this means that $p=\frac{x'-\xt}{t'-t}$, or
\begin{equation}\label{flattang}
\xt=x'-p(t'-t),
\end{equation}
although in curved space the relation $\xt(x't'|p\,x\,t)$ is only
implicitly defined by the matching of momenta.  However, the $\xt$
which achieves the momentum matching condition can alternatively be
found as the $\xt$ which minimizes $S(x't'|\xt\,p\,x\,t)$, the action
along such a path, for a given $x$, $p$ and $x'$.  This action has two
contributions, one from the initial discontinuity and one from
geodesic from $\xt$ to $x'$, giving
\begin{equation}\label{tildpspf}
S(x't'|\xt\,p\,x\,t)=p^a\tilde{y}_a+S(x't'|\xt t).
\end{equation}
Extremization with respect to $\xt$ gives the curved-space counterpart
of \eqref{flattang}, and defines the PSPF $S(x't'|p\,x\,t)$ for the
tangent interpolation recipe.  Note that while GI only defines a class
of PSPFs in curved space, with a recipe for propagating the momentum
needed to pick out a single PSPF, the TI procedure defines one unique
path and thus only one PSPF.

	In flat space, it is easy to verify that the TI prescription
leads to the PSPF \eqref{flatpspf}, by substituting \eqref{flattang}
into the flat-space equivalent of \eqref{tildpspf},
\begin{equation}
S(x't'|\xt\,p\,x\,t)=p\cdot(\xt-x)+\frac{1}{2}\frac{(x'-\xt)^2}{t'-t}.
\end{equation}
The first correction term leads to the form
(\ref{geodpspf},\ref{Glambda}) with the value $\lambda=-1$.  This
means, from the results of the previous section, that it satisfies
IOF, at least to the first non-trivial order in the distances
involved.

	In fact, the TI PSPF satisfies the involution property to all
orders, because the TI prescription obeys IOP, even over finite
intervals.  IOP can be verified for flat space as follows
(Fig.~\ref{fig:TI}b): Since the sum of the principal functions is
given by \eqref{flatpfs}, the values of $x'$ and $p'$ which extremize
it are again given by \eqref{flatinvo}.  The auxiliary point $\xt'$
for the second interval is given, analogously to \eqref{flattang}, by
$\xt'=x''-p'(t'-t)=x'$; since $\xt'$ and $x'$ co\"{\i}ncide, there is
no discontinuity in the path from $x'$ to $x''$, which is just a
classical phase space path.  The point $\xt=x'-p(t'-t)=x''-p(t''-t)$
to which the initial jump at time $t$ is made is in the same location
whether the final condition is taken to be $x''$ at $t''$ or $x'$
[given by \eqref{flatinvo}] at $t'$.

\section{CONCLUSIONS AND OUTLOOK}

	We have shown that the involution property described in
Sec.~\ref{sec:invo} is obeyed by the phase space skeletonization
procedure which we defined in Sec.~\ref{sec:TI} and named tangent
interpolation.  According to prior results, summarized in
Sec.~\ref{sec:skel}, using this skeletonization rule to construct a
phase space path integral will induce a particular measure on the
corresponding configuration space path integral.  In this way,
involution singles out a preferred realization of the skeletonized
path integral.  It is one which leads to a Schr\"{o}dinger equation
whose Hamiltonian differs from the simplest form, proportional to the
covariant Laplacian, by a term $\frac{\hbar^2}{3}R$ proportional to
the scalar curvature of the spatial manifold.  Understanding the
implications of this result rests on two as yet open questions.
First, is there any physical explanarion for the modified Hamiltonian,
and second, what is the significance of the involution property for
the path integral prescription?  Involution of the Hamilton principal
function enables composition of propagators constructed via the
configuration space path integral, for any choice of measure.  Perhaps
involution of the phase space principal function is related to
composition of path integrals defining momentum space propagators.
Further investigation is called for.

\section*{ACKNOWLEDGMENTS}

The author wishes to thank D.~A.~Craig, F.~Dowker, K.~V.~Kucha\v{r} and
S.~F.~Ross for helpful discussions.  This work was supported by NSF
grant PHY-9507719.

\end{document}